\documentclass[aps,twocolumn]{revtex4}

\begin{document}

\def\bb{\begin{equation}}
\def\ee{\end{equation}}

\title{Correlation, excitation, and relaxation of the antiferromagnetic
nanoparticle}
\author{J.D. Lee}
\affiliation{W.M. Keck Laboratories, California Institute of Technology,
Pasadena, CA 91125}
\date{\today}

\begin{abstract}
We study the dynamics of spins in the canted antiferromagnetic 
hematite nanoparticles
$\alpha$-Fe$_2$O$_3$. The system includes an antiferromagnetic exchange ($J$)
between two sublattices as well as the potential dominated by a uniaxial
anisotropy and dissipates through the thermally triggered spin-phonon
mechanism. The exchange $J$ is found to reduce the superparamagnetic
relaxation and drive the incoherent relaxation of transverse spins
in the low temperature. Dynamical properties are semi-quantitatively 
understood within a role of the exchange interaction $J$ and
agree well with recent inelastic neutron scattering experiments.
\end{abstract}


\maketitle

Magnetic properties of the fine particles of a size $\sim {\cal O}(10)$ nm
are interesting and  currently attracting much attention.
In the system, the most relevant energy scale tends to be the magnetic 
anisotropic energy, easily comparable to the thermal energy $T$, 
so that the magnetization (or {\it superspin}) is subject to considerable 
thermal fluctuation, called the {\it superparamagnetic 
relaxation}\cite{Neel}. Phenomena of such magnetic fluctuations 
have the potential importance in the fundamental research 
as well as the technological application\cite{Dormann}.

Hematite, $\alpha$-Fe$_2$O$_3$, is a mineral of the corundum (rhombohedral)
structure at room temperature, characterized by the rhombohedral
axis [111] and the basal (111) plane. Spins in hematite are sightly canted 
away from antiferromagnetic (AFM) orientation on the basal plane, giving place 
to the weakly ferromagnetic (FM) state, between the Morin temperature
$T_M\approx 260$ K and the N\'{e}el temperature $T_N\approx 956$ K, but
spins are oriented along the [111] axis and perfectly antiferromagnetic
below $T_M$. Magnetic properties are found to depend on the particle size
and $T_M$ is less than 5 K in typical nanocrystalline 
hematites\cite{Schroeer}, where several kinds of magnetic anisotropy 
energies, not present or much smaller in the bulk,
play important roles. In nanocrystalline hematites, the magnetization
oscillates due to superparamagnetic relaxation along the easy 
axis and collective magnetic excitations in vicinity of the easy 
axis\cite{Dormann}.

Many studies of the magnetic fluctuations in the nanoparticle hematite
are carried out using M\"{o}ssbauer spectroscopy or 
measuring the magnetic susceptibility.
Morup and his coworkers have used M\"{o}ssbauer spectroscopy 
and studied the energy scale of the superparamagnetic relaxation
and anisotropic barrier\cite{Bodker}. They have also explored effects of
the inter-particle interaction in strongly interacting 
nanoparticles (e.g., uncoated dry particles or a form of mixture
such as $\alpha$-Fe$_2$O$_3$+CoO or ${}^{57}$Fe-NiO+CoO)
and found the absence of fast superparamagnetic 
relaxation, i.e. the slowing down of the relaxation\cite{Hansen00}.
By the way, the magnetic relaxations in the system are argued to occur
highly fast, being almost the ultrafast process (in the time scale
of $\lesssim{\cal O}(10^{-12})$ seconds or in the frequency scale
of THz range), and the M\"{o}ssbauer spectroscopy
or susceptibility measurements are then found to be slow 
probes for the relevant magnetic dynamics. By the same group and coworkers,
due to those reasons, the inelastic neutron scattering experiment 
is suggested as one of valuable complements\cite{Hansen97}.

In this paper, we investigate the detailed dynamics of the magnetization
in the AFM nanoparticle system. The system incorporates
the AFM exchange $J$ between two sublattices
as well as the potential by the anisotropy and undergoes 
the spin-relaxation through the spin-phonon mechanism. 
A main goal of the paper is to understand the respective dynamics 
and relaxation 
of both the superparamagnetic fluctuations and the collective magnetic 
excitations with respect to the temperature $T$. Prior to the present work,
theoretical studies have analytically shown that the superparamagnetic 
relaxation follows the simple Arrhenius behavior of
$\sim \Gamma_0(T) e^{-E_B/T}$ with a certain form of $\Gamma_0(T)$,
where $E_B$ the energy barrier, for FM grains 
(like Mn$_{12}$O$_{12}$ molecules)\cite{Villain,Wurger1,Wurger2}.
At the very low temperatures ($T\lesssim 2$ K), however, 
it is found that the relaxation deviates from the Arrhenius law
and is dominated by a resonant tunneling between low-lying 
states\cite{Fort}. In the present work, such low temperature regimes
are not assumed.

Considering that spins in the bulk hematite are in the slightly
canted AFM (or weakly FM) state on the (111) basal plane, where
the hexagonal crystal anisotropy is dominated by the uniaxial one,
we suggest the Hamiltonian ${\cal H}_0$ as
\bb\label{Eq:1}
{\cal H}_0=K_z(S_{1z}^2+S_{2z}^2)+K_y(S_{1y}^2+S_{2y}^2)
          +J{\bf S}_1\cdot{\bf S}_2,
\ee
where ${\bf S}_1$ and ${\bf S}_2$ are the magnetization or superspin
(or simply {\it spin})
of two AFM sublattices, with a size of $S\gg 1$, 
and $K_z$ and $K_y$ are uniaxial anisotropic constants
along $\hat{z}$ ([111] direction) and in the basal plane, respectively,
and $J$ is the exchange between two sublattice magnetizations.
$K_z$ is expected to be the largest energy scale, i.e.
$K_z\gg K_y$, and $K_z\gg J$. $K_zS^2$ is noted to be a natural 
energy unit. ${\cal H}_0$ can be easily diagonalized
by taking the basis of $|S_{1z},S_{2z}\rangle$ and performing
the calculation of Clebsch-Gordan coefficients 
for the coupling term.
In addition to ${\cal H}_0$, we consider 
the spin-phonon interaction ${\cal V}$ ($={\cal V}_1+{\cal V}_2$)
raising the spin-relaxation, ${\cal V}_i=v_if$,
\bb\label{Eq:2}
v_i=g_0(S_{iz}S_{ix}+S_{iy}S_{iz})+g_1S_{ix}S_{iy}
   +\mbox{h.c.},
\ee
where $f$ denotes the lattice deformation.
The first term in Eq.(\ref{Eq:2}) gives rise to transitions
from $|m_i\rangle$ to $|m_i\pm1\rangle$, while
the second term leads to transitions to $|m_i\pm2\rangle$.

\begin{figure}
\vspace*{5.cm}
\includegraphics{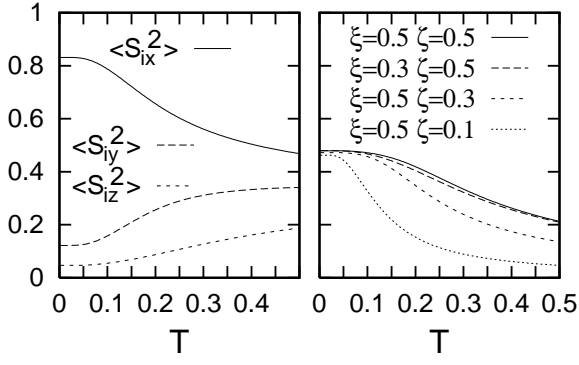}
\caption{Left panel: respective spin averages divided by $S(S+1)$ 
with respect to $T$. $\xi=0.3$ and $\zeta=0.5$.
Right panel: behaviors of $\langle ({\bf S}_1-{\bf S}_2)^2\rangle
-{\bf S}_1^2-{\bf S}_2^2$ normalized by $2S(2S+1)$ 
with respect to $T$.
}
\label{FIG1}
\end{figure}

We have interests in the relaxational dynamics of spins at the
AFM Bragg scattering, which is given by the following 
correlation function, $S_{\mu}\equiv S_{1\mu}-S_{2\mu}$,
$\mu=x,y,z$,
\bb\label{Eq:3}
\phi_{\mu\mu}(\tau)=\langle S_{\mu}(\tau)S_{\mu}\rangle
=\langle S_{\mu}e^{-i{\cal L}\tau}S_{\mu}\rangle,
\ee
where ${\cal L}$ is the Liouville operator, 
${\cal LO}=[{\cal H},{\cal O}]$, ${\cal H}={\cal H}_0+{\cal V}$.
It is the spectral function
$J_{\mu\mu}(\omega)=\int d\tau e^{i\omega\tau}\phi_{\mu\mu}(\tau)$ 
that can be compared directly with the inelastic
neutron scattering experiment. 
Up to the second order of ${\cal V}$, the correlation function 
$\phi_{\mu\mu}(\tau)$ is written as
$\phi_{\mu\mu}(\tau)
=\langle e^{i{\cal H}_0\tau}S_{\mu}e^{-i{\cal H}_0\tau}
e^{-i{\cal L}_{\cal V}\tau}S_{\mu}\rangle$, which is valid 
to the lowest order 
in the memory function\cite{Forster}. By introducing the relaxation
matrix ${\cal R}$ whose elements consist of the transition probabilities
of the Markov equation, we replace 
$e^{-i{\cal L}_{\cal V}\tau}$ by $e^{-{\cal R}\tau}$
\bb\label{Eq:4}
\phi_{\mu\mu}(\tau)
=\langle e^{i{\cal H}_0\tau}S_{\mu}e^{-i{\cal H}_0\tau}
e^{-{\cal R}\tau}S_{\mu}\rangle.
\ee
With the basis of the eigenstate of ${\cal H}_0$,
elements of the relaxation matrix ${\cal R}$ are found as follows;
${\cal R}_{nn}=\sum_{m=n+1}^{\cal N}\gamma_{n\to m}
+\sum_{m=0}^n e^{-(E_m-E_n)/T}\gamma_{m\to n}$, and
${\cal R}_{nm}=-\gamma_{m\to n}$ for $n>m$, and
${\cal R}_{nm}=-e^{-(E_n-E_m)/T}\gamma_{n\to m}$ for $n<m$,
where ${\cal N}(=(2S+1)^2)$ is the dimension of the matrix and 
finally $\gamma_{n\to m}$ is defined by\cite{Villain,Wurger1}
$$
\gamma_{n\to m}=\alpha\frac{|\langle n|v|m\rangle|^2}
{1-e^{-(E_n-E_m)/T}}(E_n-E_m)^3.
$$
\begin{figure}
\vspace*{14.cm}
\includegraphics{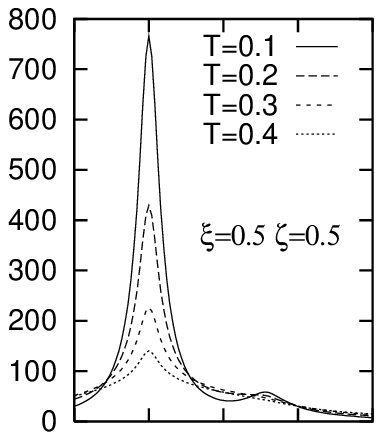}
\caption{Left panel: total spectra $J(\omega)$, a sum of
$J_{xx}(\omega)$, $J_{yy}(\omega)$, and $J_{zz}(\omega)$
for given parameter sets
with respect to $T$. Right panel: transverse spectra $J_{yy}(\omega)$
and $J_{zz}(\omega)$ and their sum $J_{\perp}(\omega)$.
The unit of $\omega$ is $2K_zS$.
}
\label{FIG2}
\end{figure}
We assume $E_n>E_m$ for $n<m$. ${\cal R}$ is in fact 
a generalization of Orbach's process\cite{Abragam} to a spin cascade.
For a coupling to elastic waves, $\gamma_{n\to m}$ should be
$\propto (E_n-E_m)^3$ from the line shape function.
It should be noted that the eigenstate of the 
relaxation matrix ${\cal R}$ does not construct an orthonormal basis set
because ${\cal R}$ is not a (real) symmetric matrix\cite{Wurger_comment}.
We then need an orthogonalization process of nonorthogonal basis,
i.e. like the Gram-Schmidt orthogonalization\cite{Arfken}. 
Putting the eigenstate of ${\cal R}$ with the relaxation eigenvalue
of $\lambda_i$ as $|p_i\rangle$,
the orthogonalized state $|\varphi_j\rangle$ is,
$|\varphi_j\rangle=C_j^j\left[|p_j\rangle-\sum_{l=0}^{j-1}
\sum_{i=0}^l C_l^i\langle p_j|\varphi_l\rangle|p_i\rangle\right]$
with $|\varphi_0\rangle=|p_0\rangle$, $C_j^j=1/\sqrt{1-\sum_{i=0}^{j-1}
|\langle p_j|\varphi_i\rangle|^2}$, and $C_l^i=0$ for $i>l$, and 
we find $C_j^i=-C_j^j\sum_{l=i}^{j-1}C_l^i\langle p_j|\varphi_l\rangle$
for $i<j$. Now the correlation function of $\phi_{\mu\mu}(\tau)$
is expressed as
\begin{eqnarray}\label{Eq:5}
\phi_{\mu\mu}(\tau)&=&\sum_n\sum_m
e^{-\beta E_n}e^{i(E_n-E_m)\tau}\langle n|S_{\mu}|m\rangle
\\ \nonumber
&\times&\sum_j\sum_{i=0}^j C_j^i e^{-\lambda_i\tau}
\langle m|p_i\rangle\langle\varphi_j|S_{\mu}|n\rangle.
\end{eqnarray}

In the left panel of Fig.\ref{FIG1}, it is shown 
that $\hat{x}$ should be the direction of the easy axis, while,
in the right panel, the temperature variation of AFM correlation strengths 
are given for $\xi$'s and $\zeta$'s ($\xi\equiv K_y/K_z$ and $\zeta\equiv 
J/K_z$). Energy quantities
including $T$ are given by a unit of $K_zS^2$, but the excitation 
energy $\omega$ (also $\Delta_y$ and $\Delta_z$ given later)
should be scaled by $2K_zS$ rather than $K_zS^2$
from $K_zS^2-K_z(S-1)^2\approx 2K_zS$.
At low temperatures ($T\ll K_zS^2$), there are two distinctly different 
types of dynamics, namely the superparamagnetic fluctuation 
$J_{\parallel}(\omega)$ ($=J_{xx}(\omega)$) 
and the collective magnetic excitation $J_{\perp}(\omega)$
($=J_{yy}(\omega)+J_{zz}(\omega)$). 

\begin{figure}
\vspace*{5.cm}
\includegraphics{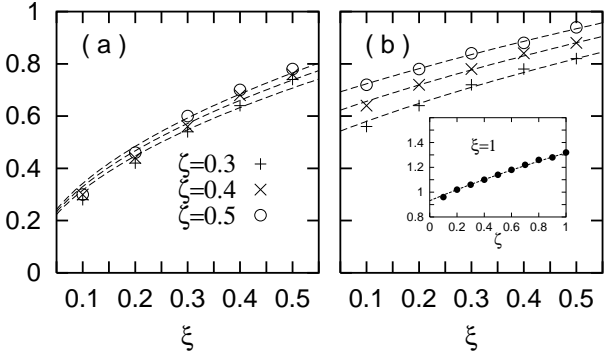}
\caption{(a) the excitation energy of $S_y$ ($\Delta_y$).
Dashed lines are fittings from
$\propto\sqrt{\xi\zeta}$. (b) the excitation energy of $S_z$ ($\Delta_z$),
$\propto\sqrt{\xi+\zeta}$. The legend in (a) is applied for (b), too.
The inset shows the transverse excitation energy ($\Delta_{\perp}
=\Delta_y=\Delta_z$) when $\xi=1$, nicely fitted with $\propto\sqrt{\zeta+1}$.
Results are evaluated at $T=0.1$. The unit of both $\Delta_y$
and $\Delta_z$ is $2K_zS$.
}
\label{FIG3}
\end{figure}

Collective magnetic excitations lead to 
the inelastic scattering process. The relative intensity of 
$J_{yy}(\omega)$ and $J_{zz}(\omega)$ depends on the ratio of
two different spin-dissipation channels in Eq.(\ref{Eq:2}). 
We take the ratio $g_1/g_0=0.5$ because 
transitions of $\delta m_i=1$ are considered to occur more probably than
$\delta m_i=2$ and $\alpha g_0^2=1\times 10^{-3}$. Under the ratio,
it is found that $J_{\perp}(\omega)$ is governed by $J_{yy}(\omega)$,
$J_{\perp}(\omega)\approx J_{yy}(\omega)$.
For three sets of parameters, we provide
the spectral functions in Fig.\ref{FIG2}. We note that two 
important features characteristically develop with respect to
$\xi$, $\zeta$, and $T$; one is the collective excitation energy and 
the other is the relaxation of spins. By comparing panels in Fig.\ref{FIG2},
it is found that the excitation energies of $S_y$ ($\Delta_y$) and 
$S_z$ ($\Delta_z$) behave differently. 
In Fig.\ref{FIG3}, the collective excitation energies
for $\xi$ and $\zeta$ are given. We find that 
$\Delta_y$ ($\approx\Delta_{\perp}$) is scaled 
roughly by $\sim\sqrt{\xi(\xi+\zeta)}$, 
that is, $\Delta_y\propto\sqrt{\xi\zeta}$
for $\xi\ll\zeta$ and $\Delta_y\propto\xi$ for $\zeta\sim 0$.
This is consistent with the result of
inelastic mode energy of a pure antiferromagnet\cite{Hansen97,Lindgard}.
On the other hand, $\Delta_z$ is by
$\sim \sqrt{\xi+\zeta}$. Especially, by the symmetry, for a case of $\xi=1$,
$\Delta_y$ should be same as $\Delta_z$,
$\Delta_{\perp}=\Delta_y=\Delta_z\propto\sqrt{\zeta+1}$,
while, for $\xi\to 0$, $\Delta_y$ goes to 0 
by $J_{yy}(\omega)=J_{xx}(\omega)$.
\begin{figure}
\vspace*{6.8cm}
\includegraphics{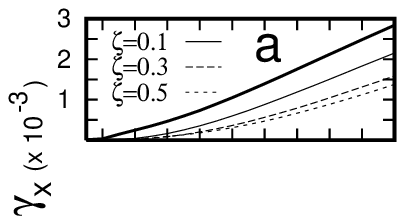}
\caption{Behaviors of spin relaxation constant $\gamma_x(T)$, $\gamma_y(T)$,
and $\gamma_z(T)$ for $\xi$'s and $\zeta$'s with respect to $T$.
Figures (a) and (b) are for $\xi=0.5$ and (c) and (d) are for $\zeta=0.5$.
Thick solid lines in (a) and (b) are for $\zeta=0$.
The legend in (a) is also for (b) and that in (c) also for (d),
respectively. (e) gives the energy barrier out of the Arrhenius law
$\gamma_x(T)\sim e^{-E_B/T}$ with respect to $\zeta$
(the line is a guide for the eye).
}
\label{FIG4}
\end{figure}
Now, let us consider the relaxation of the longitudinal
and transverse components of spins, which is in principle understood 
from the broadness of curves in Fig.\ref{FIG2}. 
Instead of directly measuring the broadness, however, for the systematic 
investigation of the relaxation, we define and propose a useful renormalized 
dissipation parameter $\gamma_{nm}^{\mu}$; 
$
\sum_j\sum_{i=0}^j C_j^i e^{-\lambda_i\tau}
\langle m|p_i\rangle\langle\varphi_j|S_{\mu}|n\rangle
\approx e^{-\gamma_{nm}^{\mu}\tau}\langle m|S_{\mu}|n\rangle,
$
where $\gamma_{nm}^{\mu}$ is found to describe the relaxation of the system
at least in an approximate sense.
It is evident that the relaxation of the system is mainly
governed by some smallest eigenvalues of $\lambda_i$'s and it is then
essential to account for the long-time behavior
after long enough $\tau$ has elapsed.
Due to the reason, we obtain $\gamma_{nm}^{\mu}$ by
integrating both sides from $\tau=0$ to $\tau=\infty$,
$\gamma_{nm}^{\mu}\approx\langle m|S_{\mu}|n\rangle/\left[
\sum_j\sum_{i=0}^j\left\{C_j^i 
\langle m|p_i\rangle\langle\varphi_j|S_{\mu}|n\rangle/
(\lambda_i+\delta)\right\}\right]-\delta$,
where the finite dissipation $\delta=2.5\times 10^{-4} K_zS^2$ 
has been added for the stable calculation. $\delta$ can be understood 
corresponding to the finite energy resolution in an actual
experiment\cite{resolution}. With $\gamma_{nm}^{\mu}$, the
time-dependent spectral function $\phi_{\mu\mu}(\tau)$ is 
$\phi_{\mu\mu}(\tau)=\sum_n\sum_m e^{-\beta E_n}e^{i(E_n-E_m
+i\gamma_{nm}^{\mu})\tau}|\langle n|S_{\mu}|m\rangle|^2$, from which 
$\gamma_{\mu}$ is further defined as the smallest value of $\gamma_{nm}^{\mu}$
at the peak position of the spectral function of $S_{\mu}$.
Behaviors of $\gamma_{\mu}(T)$ have essential
importances in that the magnetic relaxation of spins and related
dynamics of energy dissipation can be well understood through them.
We have two distinguishable kinds 
of the spin relaxation, as shown in Fig.\ref{FIG4}; one is the 
superparamagnetic relaxation along the easy axis ($\gamma_x(T)$; 
Figs.\ref{FIG4}(a) and (c)) 
and the other is the relaxation by the collective excitations
transverse to the easy axis ($\gamma_y(T)$ or $\gamma_z(T)$;
Figs.\ref{FIG4}(b) and (d)).
Because $\gamma_z(T)$ is always larger than $\gamma_y(T)$, 
consistent with $J_{yy}(\omega)\gg J_{zz}(\omega)$, it is reasonable
to put $\gamma_{\perp}(T)=\gamma_y(T)$. Like cases of FM grains, 
the superparamagnetic relaxation still follows the simple Arrhenius 
behavior as a first approximation, even if the relaxation has 
two-dimensional nature. However, the energy barrier  
depends on $\xi$ and $\zeta$ and does not have so simple a form.
From Fig.\ref{FIG4}(a), it is evident 
that the AFM exchange $J$ reduces the superparamagnetic relaxation,
in other words, the energy barrier increases with 
$J$ as shown in Fig.\ref{FIG4}(e). This finding
is consistent with the claim of the absence of the fast superparamagnetic
relaxation in the interacting nanoparticles through
the M\"{o}sbauer spectroscopy. Without the
exchange, the energy barrier approximately follows
$\sim K_yS^2$, which implies $\gamma_x\propto e^{-\xi/T}$;
$\gamma_x\propto e^{-0.46/T}$ for $\xi=0.5$ and $\zeta=0$. 
Figures \ref{FIG4}(b) and (d) give the transverse relaxation,
quite different from the superparamagnetic relaxation.
In the low temperatures, to increase the temperature 
reduces $\gamma_{\perp}(T)$.
$\gamma_{\perp}(T)$ has then a minimum and slowly increases with $T$.
In such a sense, the transverse responses of spins are found incoherent
in the low temperatures. We conclude that the incoherence is driven by
the AFM spin exchange $J$ because $\gamma_{\perp}(T)$ is always coherent
for $J=0$ (a thick line in Fig.\ref{FIG4}(b)).
The incoherent behaviors of $\gamma_{\perp}(T)$ in the low temperatures
has been actually observed in the inelastic neutron scattering experiment
for $\alpha$-Fe$_2$O$_3$\cite{Hansen97}. 

Here we can estimate
the system parameters $K_zS^2$, $K_yS^2$, and $JS^2$ from
the scaled dynamical quantities\cite{scaling}. Their estimations have been 
done with appreciable error ranges from temperature fits in various ways by
Morup and his coworkers\cite{Hansen97}. A typical one of them is 
$K_yS^2\sim 500$ K (i.e. $\sim 42$ meV) from Arrhenius behavior of
the superparamagnetic relaxation, but it is found from Fig.\ref{FIG4}(e)
that the value may have been overestimated almost by 60 $\%$.
Thus, we estimate $K_yS^2\sim 300$ K ($\sim 25$ meV), while
the value of $K_zS^2$ is more uncertain because its value is not known
for the nanoscopic system, but of a same order of magnitude with the bulk case.
So we take $K_zS^2\sim 1000$ K ($\sim 83$ meV), that is, we have
$\xi=0.3$. On the other hand, the unit excitation energy 
$2K_zS$ can be found $\sim 0.5$ meV to be consistent with the 
experiment. The inelastic energy mode ($\Delta_{\perp}$) is then
$\sim 0.3$ meV at $T=0.1$ ($\sim 100$ K). 
The value of $JS^2$ can be estimated as 300-500 K
from the transverse excitation energy $\Delta_{\perp}$ in Fig.\ref{FIG3} and 
from the relaxational behaviors of spins in Fig.\ref{FIG4}.
Based on the estimations, $\gamma_x(T)$ is about 
${\cal O}(10^{-3})$ meV and $\gamma_{\perp}(T)$ about ${\cal O}(10^{-1})$ meV
in the temperatures $T\lesssim 0.2$ (i.e. $T\lesssim 200$ K).
We note $\gamma_{\perp}(T)$ is larger than $\gamma_x(T)$ by ${\cal O}(10^2)$,
which is consistent with the actual 
situation of the experiment\cite{Hansen97}.

In summary, we have investigated the dynamics of spins in 
the antiferromagnetic hematite nanoparticle $\alpha$-Fe$_2$O$_3$
with the explicit AFM exchange $J$. The exchange $J$ has  been found to
reduce the superparamagnetic relaxation and also induce the incoherent
oscillation of transverse spins in the low temperature. 
The spin responses are well understood within the model 
and a semi-quantitative estimation of the relevant parameters 
has been done through
a comparison with the recent inelastic neutron scattering experiments.

This work was supported
by the U.S. Department of Energy under contract DE-FG03-01ER45950.


\begin{references}

\bibitem{Neel} L. N\'{e}el, Ann. Geophys. {\bf 5}, 99 (1949);
               W.F. Brown Jr., Phys. Rev. {\bf 130}, 1677 (1963).
\bibitem{Dormann} For a review, see J.L. Dormann {\it et al.},
                  Adv. Chem. Phys. {\bf 98}, 283 (1997).
\bibitem{Schroeer} D. Schroeer and R.C. Nininger, Jr., Phys. Rev. Lett.
                   {\bf 19}, 632 (1967); N. Yamamoto, J. Phys. Soc. Jpn.
                   {\bf 24}, 23 (1968); N. Amin and S. Arajs,
                   Phys. Rev. B {\bf 35}, 4810 (1987).
\bibitem{Bodker} F. Bodker {\it et al.},
                 Phys. Rev. B {\bf 61}, 6826 (2000).
\bibitem{Hansen00} M. F. Hansen {\it et al.}, Phys. Rev. B
                   {\bf 62}, 1124 (2000); C. Frandsen and S. Morup,
                   J. Magn. Magn. Mater. {\bf 266}, 36 (2003).
\bibitem{Hansen97} M.F. Hansen {\it et al.}, Phys. Rev. Lett. {\bf 79}, 4910
                   (1997); M.F. Hansen {\it et al.}, J. Magn. Magn. Mater.
                   {\bf 221}, 10 (2000); S.N. Clausen {\it et al.},
                   J. Magn. Magn. Mater. {\bf 266}, 68 (2003).
\bibitem{Villain} J. Villain {\it et al.}, 
                  Europhys. Lett. {\bf 27}, 159 (1994)
\bibitem{Wurger1} A. W\"{u}rger, Phys. Rev. Lett. {\bf 81}, 212 (1998).
\bibitem{Wurger2} A. W\"{u}rger, Europhys. Lett. {\bf 44}, 103 (1998).
\bibitem{Fort} A. Fort {\it et al.}, Phys. Rev. Lett. {\bf 80}, 612 (1998).
\bibitem{Forster} D. Forster, {\it Hydrodynamic Fluctuations, Broken
                  Symmetry, and Correlation Functions} (Benjamin, 1975).
\bibitem{Abragam} A. Abragam and B. Bleaney, {\it Electron Paramagnetic 
                  Resonance of Transition Ions} (Clarendon
                  Press, 1970).
\bibitem{Wurger_comment} Reference \cite{Wurger1} argues the superparamagnetic
                 relaxation for FM grains should be determined 
                 by the smallest finite eigenvalue $\lambda_0$ of ${\cal R}$.
                 But it is rather unclear in that the coefficient of
                 $e^{-\lambda_0\tau}$ is not necessarily positive-definite
                 due to the nonorthonormality of eigentstates of ${\cal R}$.
\bibitem{Arfken} G. Arfken, {\it Mathematical Methods for Physicists}
                 (Academic Press, 1985).
\bibitem{Lindgard} P.A. Lindgard {\it et al.},
                   J. Phys. C: Solid State Phys. {\bf 8}, 1059 (1075).
\bibitem{resolution} By our estimation ($K_zS^2\sim 83$ meV),
                     a finite dissipation (or broadness)
                     $\delta=2.5\times 10^{-4}$ is corresponding to
                     $\sim 20$ $\mu$eV, which is the same order of
                     the energy resolution adopted in Ref.\cite{Hansen97}.
\bibitem{scaling} From estimated $K_zS^2$ and $2K_zS$,
                  $S\sim {\cal O}(10^2)$ is easily found.
                  But actual evaluations are done for a smaller $S=10$
                  (${\cal N}=441$) due to practical limitations.
                  Nevertheless, thanks to scaling behaviors 
                  (all the relevant quantities 
                  are scaled by either $K_zS^2$ or $2K_zS$;
                  that is, results of Figs.\ref{FIG1}-\ref{FIG4} hardly
                  depend on a value of $S(\gtrsim 10)$ except for 
                  the absolute intensity of the spectra $J_{\mu\mu}(\omega)$'s
                  in Fig.\ref{FIG2}), reliable estimations are found possible.
                  


%

\end{references}
\end{document}